\documentclass{ws-ijmpcs}

\begin{document}

\markboth{A. Saleev \& N. Nikolaev \& F. Rathmann}
{Systematics limitations in the EDM searches.}

%
\catchline{}{}{}{}{}
%

\title{Studies of systematic limitations in the EDM searches at storage rings}

\author{Artem Saleev}

\address{
Institut f\"ur Kernphysik, Forschungszentrum J\"ulich\\
J\"ulich, 52425, Germany,\\
 and Samara State Aerospace University\\ Samara, 443086,  Russia\\
a.saleev@fz-juelich.de}

\author{Nikolai Nikolaev}

\address{L. D. Landau Institute for Theoretical Physics\\
Moscow region, Chernogolovka, 142432, Russia\\
nikolaev@itp.ac.ru}

\author{Frank Rathmann}

\address{Institut f\"ur Kernphysik and J\"ulich Center for Hadron Physics, Forschungszentrum J\"ulich\\
J\"ulich, 52425, Germany\\
f.rathmann@fz-juelich.de}

\author{On behalf of JEDI collaboration}

\maketitle

\begin{history}
\received{Day Month Year}
\revised{Day Month Year}
\published{Day Month Year}
\end{history}

\begin{abstract}
Searches of the electric dipole moment (EDM) at a pure magnetic ring, like COSY, encounter strong background coming from magnetic dipole moment (MDM). The most troubling issue is the MDM spin rotation in the so-called imperfection, radial and longitudinal, B-fields.
To study the systematic effects of the imperfection fields at COSY we proposed the original method which makes use of the two static solenoids acting as artificial imperfections. Perturbation of the spin tune caused by the spin kicks in the solenoids probes the systematic effect of cumulative spin rotation in the imperfection fields all over the ring. The spin tune is one of the most precise quantities measured presently at COSY at $10^{-10}$ level. The method has been successfully tested in September 2014 run at COSY, unravelling strength of spin kicks in the ring's imperfection fields at the level of $10^{-3} rad$.

\end{abstract}

\ccode{PACS numbers: 13.40.Em, 11.30.Er, 29.20.Dh, 29.27.Hj}

\section{EDM searches at all-magnetic storage rings}	
Spin motion in the pure magnetic ring with a vertical guiding field $\vec{B}$ is governed by T-BMT equation\cite{jpap,llmnf}
\begin{eqnarray}
\frac{d\vec{S}}{dt}&=&\vec{\Omega}\times\vec{S} \nonumber\\
\vec{\Omega}&=& -\frac{e}{m}\left\{G\vec{B} + \eta\vec{\beta}\times\vec{B}\right\}
\end{eqnarray}
where $G$ is a magnetic anomaly and $\eta=\frac{dm}{e}$ has a meaning of the ratio of the EDM, $d$, to the MDM. The generic signal of the EDM is a rotation of the spin in an electric field. The EDM interacts with the motional electric field, which tilts the stable spin axis,
\begin{equation}
\vec{\Omega}= 2\pi f_R \frac{G\gamma}{\cos\xi}\left\{\vec{e}_x\sin\xi + \vec{e}_y\cos\xi\right\}\, ,
\end{equation}
where $\tan\xi=\frac{\eta\beta}{G}$ and $f_R$ is cyclotron frequency.

In a realistic ring, the MDM spin rotation in the imperfection fields all over the ring adds to the EDM rotation in non-trivial way. For the particle on the closed orbit, the imperfection fields induce the spin kicks which repeat after each turn. The product of consecutive spin rotations all over the ring for one turn gives a rotation around spin closed orbit $\vec{c}$ with spin tune $Q_S$. As a result, spin rotation, given at some point of the ring after $k=f_R t$ turns is defined as:
\begin{equation}\label{eq:otm}
\Psi(k)=\left(e^{-i\vec{\sigma}\cdot\vec{c}\pi Q_S}\right)^k\Psi(0)
\end{equation}
where $\Psi$ is a two-component spinor and the $i$-th component of the spin vector is given by
\begin{equation}
S_i=\langle\Psi|\sigma_i|\Psi\rangle
\end{equation}
$\vec{\sigma}$ is a vector of Pauli matricies, 
\begin{equation}
\vec{\sigma}\cdot\vec{c}=c_1\sigma_1+c_2\sigma_2+c_3\sigma_3
\end{equation}
where $c_{1,2,3}$ are directional cosines for the vector of spin closed orbit. In case of non-zero EDM and no imperfections present in the ring, $c_1=\sin\xi$, $c_2=\cos\xi$, $c_3=0$, and spin tune $Q_S= \frac{G\gamma}{\cos\xi}$.

To study how the systematic effects from the imperfection fields at COSY induce changes in the spin closed orbit we proposed an original method in which an artificial imperfection fields are manipulated and then specific changes in the value of spin tune are observed.

\subsection{JEDI  September 2014 run}

At September 2014 run, conducted by JEDI (J\"ulich Electric Dipole moment Investigations) collaboration at COSY, two solenoids from the compensation magnets of the electron coolers have been used as static spin rotators which produce artificial imperfection spin kicks.

\begin{figure}[pb]
\centerline{\includegraphics[width=6.7cm]{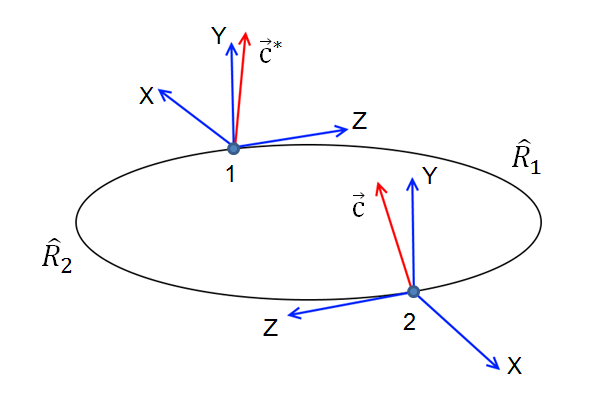}}
\vspace*{8pt}
\caption{Scheme of the ring cofiguration. The solenoids are located at points 1 and 2. Vectors  $\vec{c}$ and $\vec{c}^{\,*}$ define direction of spin closed orbit when solenoids are turned off. \label{rsketch}}
\end{figure}

Let us consider the case when the solenoids are switched off.  They are located at straight sections of the ring opposite to each other (see fig. $\ref{rsketch}$) and effectively split the ring in two halves. Each half is described by the corresponding spin rotation matrix, $\hat{R}_1$ and $\hat{R}_2$:
\begin{eqnarray}
\label{eq:otm1}\hat{R}_1&=&e^{-\frac{i}{2}\vec{\sigma}\cdot\vec{m}~\pi Q_1}\\
\label{eq:otm2}\hat{R}_2&=&e^{-\frac{i}{2}\vec{\sigma}\cdot\vec{n}~\pi Q_2}
\end{eqnarray}

Then the total spin rotation matrix at the point downstream the first solenoid:
\begin{equation}\label{eq:rax}
\hat{R}=\hat{R}_1\hat{R}_2=\exp\left\{-\frac{i}{2}\vec{\sigma}\cdot\vec{c}~2\pi Q_S\right\} ,
\end{equation}
whereas downstream the second solenoid it is different:
\begin{equation}\label{eq:raxx}
\hat{R}=\hat{R}_2\hat{R}_1=\exp\left\{-\frac{i}{2}\vec{\sigma}\cdot\vec{c}^{\,*}2\pi Q_S\right\}.
\end{equation}

The rotation axis, $\vec{c}^{\,*}=c^*_1\vec{e}_x + c^*_2\vec{e}_y + c^*_3\vec{e}_z$, is different from that one in ``($\ref{eq:rax}$)", $\vec{c}^{\,*}\neq\vec{c}$ due to non-commuting property of the rotations. The spin tune, $\nu_s=Q_S$, does not depend on the order of rotations and it is the same in both cases.

The biggest contribution to the spin rotation in each half of the ring comes from the vertical guiding field of the main dipoles in the arcs. Vertical shifts of the quadrupoles, inclinations of the dipole fields and other misalignments of magnets produce small horizontal imperfection fields. Spin kicks in the imperfection fields slightly perturb spin rotation in the vertical guiding field, and give rise to the $m_{1,3}$ and $n_{1,3}$ in Eqs. $\ref{eq:otm1},\ref{eq:otm2}$. But misalignments differ in each half of the ring, so the corresponding spin rotations for the two halves are slightly different as well, $\hat{R}_1\approx\hat{R}_2$. In the first order, 
\begin{eqnarray}
Q_1&\cong& Q_2\approx G\gamma\\
m_2&\cong& n_2\approx 1.
\end{eqnarray}

When the e-cooler's solenoids are switched on, they produce spin kicks around the longitudinal axis,
\begin{equation}\label{eq:skick}
\chi_i= (1+G)\frac{\int B_i dl}{B\rho}
\end{equation}
where $B\rho$ represents magnetic rigidity of the ring. The magnitude of $\int B dl$ has linear dependence with respect to the applied current $J_1$ or $J_2$ in each solenoid. It is calculated by using corresponding calibration factors $f_1$ or $f_2$: 
\begin{equation}
\int B_{i}dl=f_{i}J_{i}
\end{equation}
The spin kicks in the solenoids produce perturbation of the spin tune in the ring, $\nu_s=Q_S+\Delta\nu_s$. The total spin rotation matrix, defined downstream the first solenoid, becomes
\begin{equation}
\hat{R}=\hat{R}_2 \hat{R}_z(\chi_2) \hat{R}_1 \hat{R}_z(\chi_1)
\end{equation}

Using the relation $\cos(\pi\nu_s)=Tr[\hat{R}]/2$, perturbation of the spin tune is given by
\begin{eqnarray}\label{eq:stsh}
\cos(\pi Q_S)&-&\cos(\pi (Q_S+\Delta\nu_s))=\nonumber\\
&&(E+\cos(\pi Q_S))\sin^2\left(\frac{y_+}{2}\right)-\frac{1}{2}(c_3+c^*_3)\sin(\pi Q_S)\sin y_+ \nonumber\\
&-&(E-\cos(\pi Q_S))\sin^2\left(\frac{y_-}{2}\right)+\frac{1}{2}(c_3-c^*_3)\sin(\pi Q_S)\sin y_-
\end{eqnarray}
where the solenoid's spin kicks are defined as
\begin{equation}
y_\pm=\frac{\chi_1\pm\chi_2}{2} 
\end{equation}
and paramaeter $E$ is
\begin{equation}
E\approx\cos\frac{\pi(Q_1-Q_2)}{2}\approx 1
\end{equation}

\begin{figure}[pb]
\centerline{\includegraphics[width=12cm]{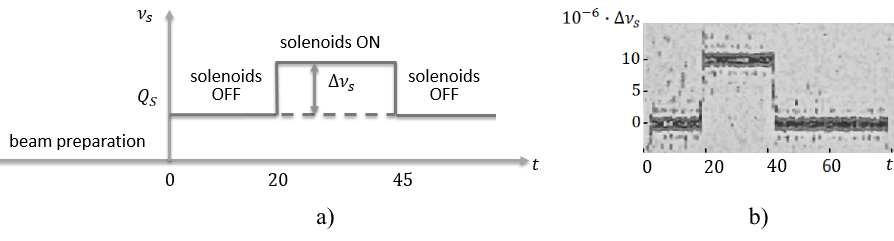}}
\vspace*{8pt}
\caption{a) Expected spin tune behaviour during each measurement. b) Typical output of the spin tune analysis software, here the spin tune shift $\Delta\nu_s\simeq10^{-5}$ has been observed between 20 and 45 seconds. \label{fstep}}
\end{figure}

The experiment consists of multiple spin tune measurements\cite{publ}. At first, the polarized deuteron beam is prepared for the spin tune measurement: after injection and acceleration to $T=270~MeV$, the beam is cooled, then bunched. Then RF-solenoid flips the initial vertical polarization into the horizontal plane by partial Froissart-Stora scan and time clock starts. All particle spins starts precessing with the spin tune $\nu_s=Q_S$ in the horizontal plane. After 20 seconds the static solenoids are switched on at specified currents $J_1$ and $J_2$. The spin tune becomes $\nu_s=Q_S+\Delta\nu_s(\chi_1,\chi_2)$. To suppress systematic effects related to the drift of the spin tune during the cycle, at 45 seconds the solenoids are switched off and $\nu_s=Q_S$ again (see fig. $\ref{fstep}$). Special data analysis software determines the spin tune shift $\Delta\nu_s$ and base spin tune $Q_S$ with the precision of $\sim10^{-10}$. To maintain a long spin coherence time of horizontal polarization, the sextupole magnets in the ring have been set up at specified settings which compensate decoherence effects from emittance and momentum spread of the beam\cite{proed}.
\begin{figure}[pb]
\centerline{\includegraphics[width=6.7cm]{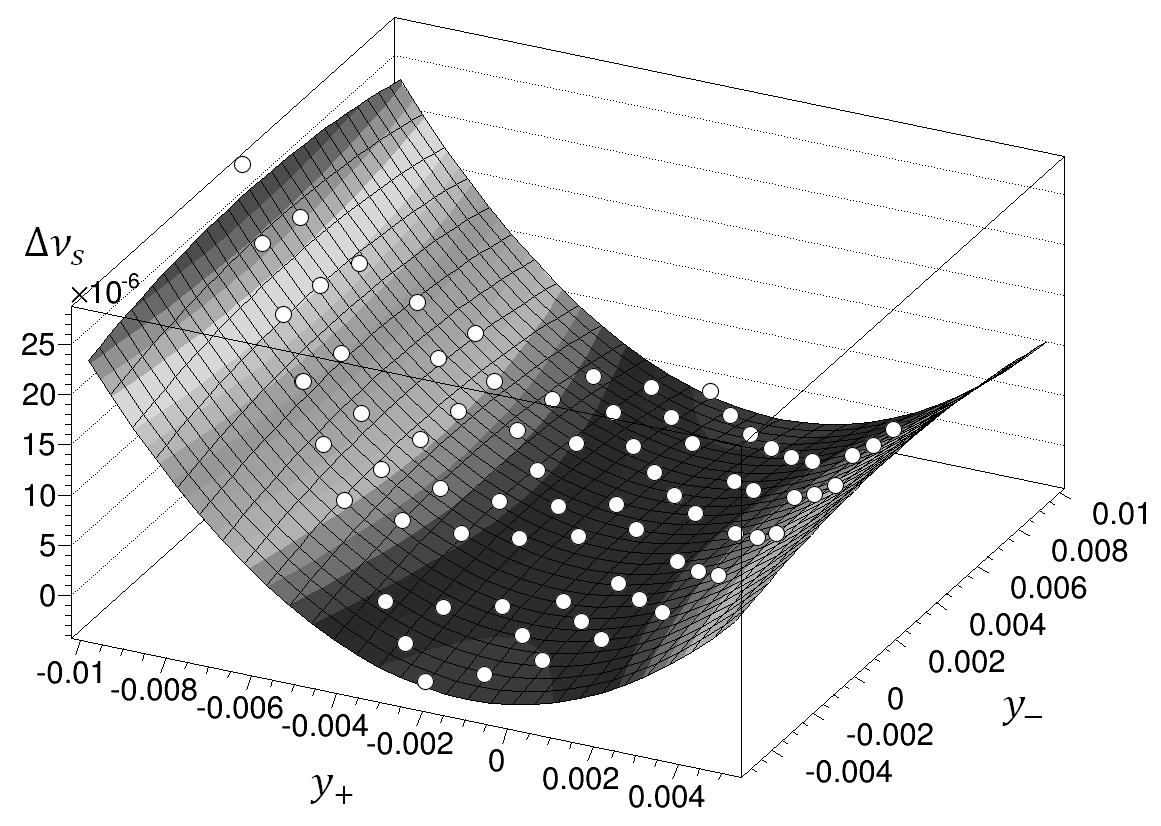}}
\vspace*{8pt}
\caption{Spin tune map. Each white dot represent single measurement of $\Delta\nu_s(y_+,y_-)$, the surface is a fit to the dataset. Error bars are smaller than the size of the symbols. Note that weak parabolic dependence in $y_-$ has negative curvature. \label{map}}
\end{figure}

The applied currents for the solenoids have been chosen such that corresponding spin kicks $\chi^i_1$ and $\chi^j_2$ form a rectangular mesh of $i\times j$ points. 

According to "Eq.$\ref{eq:stsh}$", the functional form of $\Delta\nu_s(y_+,y_-)$ is a sum of two, concave and convex, parabolas:
\begin{equation}
\Delta\nu_s(y_+,y_-)\sim\pm(y_\pm-a_\pm)^2
\end{equation}
and the extremum, the saddle point, is located at
\begin{eqnarray}
a_+&=&\sin(\pi Q_S)\frac{c_3+c^*_3}{1+\cos(\pi Q_S)} \\
a_-&=&\sin(\pi Q_S)\frac{c_3-c^*_3}{1-\cos(\pi Q_S)} 
\end{eqnarray}
where $c_3$ and $c^*_3$ are the longitudinal projections of spin closed orbit at the points 1 and 2 of the ring, as defined in "Eq. $\ref{eq:rax}$" and "Eq. $\ref{eq:raxx}$".

The procedure of the spin tune measurements with different strength of the spin kicks in static spin rotators can be referred as ``spin tune mapping". As a matter of fact, it provides a partial determination of the stable spin axis at the point of the applied spin kick, which in turn, gives a hint on the magnitude of the imperfection spin kicks in the ring.

To recover a saddle point in the spin tune map with two solenoids, the function ``($\ref{eq:stsh}$)" was fit to data points  $\Delta\nu_s(y_+,y_-)$. Fig. $\ref{map}$ shows one of the maps that has been measured during September 2014 run. The saddle point is located at $a_+=-0.00111077\pm 6.1*10^{-8}~rad$ and $a_-=-0.00244326\pm 2.05*10^{-7}~rad$, which defines $c_3=-0.00299124\pm 1.8*10^{-7}$ and $c^*_3=-0.00163653\pm 7.1*10^{-8}$. The results for $a_\pm$ are the preliminary ones, more scrutiny of the impact of possible misalignment of the solenoid axes is in order.

The method also enables to measure $c_1$ and $c^*_1$ if the static Wien filters operating as the MDM rotators with the radial rotation axis are used alongside the solenoids\cite{seb}. Part of the September 2014 run has also applied this method to studies of the systematic effects coming from the steering magnets\cite{stas}.


\section{Summary and Outlook}
The spin tune mapping emerges as a tool to determine the spin closed orbit with an unprecedented accuracy. In the present experiment only the longitudinal projection of the stable spin axis has been measured, but adding a static Wien filter next to the solenoid would enable to determine the radial projection of the stable spin axis as well. We anticipate that the spin tune mapping technique would prove most useful in a calibration of various devices to be employed in the high precision EDM searches at all magnetic rings.


\end{document}